\documentclass[fleqn,usenatbib]{mnras}

\usepackage[T1]{fontenc}

\DeclareRobustCommand{\VAN}[3]{#2}
\let\VANthebibliography\thebibliography
\def\thebibliography{\DeclareRobustCommand{\VAN}[3]{##3}\VANthebibliography}

\providecommand{\noopsort}[1]{}

\usepackage{graphicx}	
\usepackage{amsmath}	
\usepackage{amssymb}	

\usepackage{bm}

\usepackage{tikz,xcolor,hyperref}
\usepackage{graphicx}
\graphicspath{{images/}}

\definecolor{lime}{HTML}{A6CE39}
\DeclareRobustCommand{\orcidicon}{%
	\begin{tikzpicture}
	\draw[lime, fill=lime] (0,0) 
	circle [radius=0.16] 
	node[white] {{\fontfamily{qag}\selectfont \tiny ID}};
	\draw[white, fill=white] (-0.0625,0.095) 
	circle [radius=0.007];
	\end{tikzpicture}
	\hspace{-2mm}
}

\newcommand{\orcidVP}{\href{https://orcid.org/0000-0002-3031-062X}{\orcidicon}}
\newcommand{\orcidEV}{\href{https://orcid.org/0000-0003-2742-6872}{\orcidicon}}

\defcitealias{pav_ves_letter}{Paper~I}

\newcommand{\Msun}{M_\odot}

\newcommand{\trh}[1][]{t_\mathrm{rh#1}}
\newcommand{\tcc}[1][]{t_\mathrm{cc#1}}
\newcommand{\ra}[1][]{r_\mathrm{a#1}}
\newcommand{\rh}[1][]{r_\mathrm{h#1}}
\newcommand{\rt}[1][]{r_\mathrm{t#1}}

\newcommand{\disp}[1][]{\sigma_\mathrm{#1}}
\newcommand{\meq}[1][]{m_\mathrm{eq#1}}
\newcommand{\etam}[1][]{\eta_\mathrm{#1}}

\newcommand{\ft}{f_\mathrm{t}}
\newcommand{\fb}{f_\mathrm{b}}
\newcommand{\Nb}{N_\mathrm{b}}
\newcommand{\Ncm}{N_\mathrm{cm}}

\usepackage{newtxtext,newtxmath}

\title[Energy equipartition: tides, binaries, kinematics]{Evolution towards energy equipartition in star clusters: effects of the tidal field, primordial binaries, and internal velocity anisotropy}

\author[Pavl\'ik \& Vesperini]{
V\'aclav Pavl\'ik\,$^{1}$\thanks{E-mail: vpavlik@iu.edu}\orcidVP,
Enrico Vesperini\,$^{1}$\orcidEV
\\
$^{1}$Indiana University, Department of Astronomy, Swain Hall West, 727 E 3$^\text{rd}$ Street, Bloomington, IN, 47405, USA
}

\date{Accepted 2021 October 27. Received 2021 October 19; in original form 2021 August 30}

\pubyear{2021}

\begin{document}
\label{firstpage}
\pagerange{\pageref{firstpage}--\pageref{lastpage}}
\maketitle

\begin{abstract}
This paper is the second in a series investigating the evolution of star clusters towards energy equipartition (EEP). Here, we focus on the effects of the external tidal field of the host galaxy, initial anisotropy in the velocity distribution, and primordial binary star population. The results of our $N$-body simulations show that regardless of the strength of the tidal field or the fraction of primordial binaries: (i) the evolution towards EEP in the intermediate and outer regions of initially anisotropic systems is more rapid than for isotropic systems; (ii) this evolution also proceeds at different rates for the tangential and radial components of the velocity dispersion; and (iii) the outer regions of the initially isotropic systems show a tendency to evolve towards a state of `inverted' EEP in which low-mass stars have smaller velocity dispersion than high-mass stars. We also find that the clusters with primordial binaries stay even farther from EEP than systems containing only single stars.
Finally, we show that all these results also hold when the degree of EEP is calculated using quantities measured in projection as it is done in observational studies, and that our findings could be tested with current and upcoming observational data.
\end{abstract}

\begin{keywords}
globular clusters: general -- stars: binaries: general -- stars: kinematics and dynamics -- methods: numerical
\end{keywords}



\section{Introduction}
\label{sec:intro}

The evolution towards energy equipartition (EEP) -- a state of a dynamical system in which its constituents have evenly distributed kinetic energy (i.e., in a star cluster, higher-mass stars have lower velocity dispersion than lower-mass ones) -- is a direct consequence of two-body relaxation processes. However, in real clusters (and their $N$-body representations) this ideal state of complete EEP throughout the entire system never occurs because of an extremely long relaxation time of the cluster halo \citep{spitzer69,inagaki_wiyanto,spitzer}, so we may only expect EEP in the cluster core \citep[e.g.][]{merritt}. However, once the effects of relaxation and mass segregation form a core dominated by massive stars and decoupled from the surrounding lower-mass stars, the evolution towards EEP is halted (e.g.~\citealt{omegaCen_noequip,webb_vesperini_b}; see also, e.g., \citealt{libralato_hst,baldwin_hst} for recent observational studies of EEP in Galactic globular clusters).

Most theoretical and numerical investigations of the evolution towards EEP assumed the conventional dynamical picture of globular clusters (GCs) according to which these systems are non-rotating and have an isotropic velocity distribution \citep[see, e.g., the review of][and references therein]{Varri2018}. Many studies showed, however, that GCs are likely characterised by more complex kinematic properties, including anisotropy and rotation \citep[see, e.g., the review of][]{varri_mmsai}. Specifically, the radially anisotropic velocity distribution may come from either the early violent relaxation phase \citep[e.g.][]{vanAlbada,tre_ber_van,vesperini_etal14} or the long-term evolution \citep[e.g.][]{spitzer_shapiro,spitzer,giersz_heggie94,Tio_Ves_Var16}. In fact, more recent observational studies provided evidence of anisotropy in several Galactic GCs \citep[e.g.][]{watkins_hst,bellini_hstV,jindal_gaia,vasiliev_gaia}, and extensive multi-object spectroscopic studies and proper motion surveys with \textsl{HST} and \textsl{Gaia} demonstrated that many Galactic GCs also display internal rotation \citep[e.g.][]{kamann,bellini_hstV,MUSE,ferraro_etal,bianchini_gaia,sollima_gaia}.

These findings, in turn, provided new motivation for theoretical efforts to study the role of rotation in the evolution of star clusters \citep[e.g.][]{fokkerplanck_rotI,fokkerplanck_rotII,ernst_etal,hong_etal,Tio_Ves_Var16,Tio_Ves_Var17}. As for the velocity anisotropy, a number of studies explored its role for a variety of fundamental questions concerning the GCs formation history \citep[e.g.,][]{cluster_history}, their early and long-term dynamical evolution \citep{Tio_Ves_Var16,breen_var_heg,bianchini_core_collapse}, the dynamical evidence of a central intermediate-mass black hole \citep{zocchi17,imbh_anisotropy}, or differences between multiple stellar populations \citep[][]{hst_UV_legacy,Tio_Ves_Var19,cordoni_etal20,vesperini_etal21,sollima21}.

In the first paper of the series \citep[][hereafter \citetalias{pav_ves_letter}]{pav_ves_letter}, we presented novel results from a study aimed at exploring the dependence of the evolution towards EEP on the initial degree of radial anisotropy. These results showed that the evolution towards EEP depends on: (i) the degree of the initial radial anisotropy in the stellar velocities; (ii) the component of the velocity dispersion (radial or tangential); and (iii) the distance from the cluster centre where the simulations also revealed that the outer regions of an initially isotropic model evolve away from EEP, i.e., towards a state of `inverted' EEP.
In \citetalias{pav_ves_letter}, we focused on a tidally underfilling model with an initial ratio of the tidal radius to the King model truncation radius equal to 10, and no primordial binaries.
In this work, we extend our analysis to a set of star cluster models with different filling factors and also to models with primordial binaries. Furthermore, we also present our results on the evolution towards EEP using the projected velocity dispersion to offer a more direct comparison with observational quantities.

\section{Methods}
\label{sec:methods}
We present a set of $N$-body models of star clusters that were set up initially using the \citet{king_model} density profile with the central dimensionless potential $W_0 = 6$.
The modelled clusters were placed on various circular orbits in a point-like external potential that determined the initial filling factor, $\ft$, of each cluster, defined as the ratio of the truncation radius to the tidal radius, $\rt$.
Individual stellar masses were drawn from the range $0.1 \leq m / \Msun \leq 1.0$ according to the \citet{kroupa} initial mass function.

In some models, we injected a primordial population of binary stars that made up a fraction $\fb \equiv \Nb \,/\, \Ncm = 0.1$, where $\Nb$ is the number of centres of mass of the binaries and $\Ncm = 10^5$ (in all models) is the total initial number of centres of mass. Thus, the total number of stars is $N_0 = \Ncm$ in the models without binaries, and $N_0 = \Ncm+\Nb$ in the models with binaries.
The binary population has a uniform distribution in the log of the binding energy between $1$ and $30\,kT$, a thermal distribution of eccentricities \citep[$f(e)=2e$; see, e.g.,][]{heggie75} and secondary masses drawn from a uniform distribution between $0.1\,\Msun$ and the primary mass.
We note that while analysing the evolution of our models (e.g., calculating the velocity dispersion or $\meq$, see Sect.~\ref{sec:results}), all dynamically formed and primordial binaries were treated as single point masses with positions and velocities equal to those of the corresponding binary centre of mass, and masses equal to the sum of the components.

\begin{table}
	\centering
	\caption{Parameters of the models used in this work: model label, initial anisotropy radius, filling factor, and primordial binary fraction.}
	\begin{tabular}{p{2cm}p{1cm}p{1cm}p{1cm}}
		\hline
		label $^{1}$          & $\ra$     & $\ft$ & $\fb$ \\
		\hline         
		\texttt{i0.1s} $^{2}$ & $\infty$  & 0.1   & 0.0   \\
		\texttt{i0.3s}        & $\infty$  & 0.3   & 0.0   \\
		\texttt{i0.5s}        & $\infty$  & 0.5   & 0.0   \\
		\texttt{i1.0s}        & $\infty$  & 1.0   & 0.0   \\
		\texttt{a0.1s} $^{2}$ & $\rh[,0]$ & 0.1   & 0.0   \\
		\texttt{a1.0s}        & $\rh[,0]$ & 1.0   & 0.0   \\
		\texttt{i0.1b}        & $\infty$  & 0.1   & 0.1   \\
		\texttt{i1.0b}        & $\infty$  & 1.0   & 0.1   \\
		\texttt{a0.1b}        & $\rh[,0]$ & 0.1   & 0.1   \\
		\texttt{a1.0b}        & $\rh[,0]$ & 1.0   & 0.1   \\
		\hline
		\multicolumn{4}{p{6cm}}{\small $^{1}$  The label is a combination of a letter \texttt{i}/\texttt{a} for models with a radially isotropic/anisotropic initial velocity distribution, a~number representing the filling factor, and a~letter \texttt{s}/\texttt{b} indicating whether the model includes only single stars or also primordial binaries.} \\
		\multicolumn{4}{p{6cm}}{\small $^{2}$ Models presented in \citetalias{pav_ves_letter}.} \\
	\end{tabular}
	\label{tab:models}
\end{table}

The evolution of our models was followed numerically with the direct $N$-body integrator {\sc nbody6++gpu} \citep{nbody6pp}. Since we only focus on the effects of the dynamical evolution due to two-body relaxation, we did not consider the effects of stellar evolution. Throughout this paper, time is displayed in units of the initial half-mass relaxation time \citep[see, e.g.,][]{spitzer}, defined as
\begin{equation}
	\label{eq:trh}
	\trh[,0] = 0.138 N_0\,\rh[,0]^{3/2} \big/ \ln{(0.02 N_0)} \,,
\end{equation}
in H\'enon units, where $\rh[,0]$ is the initial half-mass radius of the cluster.

For the initial velocity distribution, we used the same prescription as in \citetalias{pav_ves_letter}, that is, the anisotropy between the radial and tangential components of the velocity dispersion ($\disp[rad]$ and $\disp[tan]$, respectively) is described by the Osipkov--Merritt profile \citep[see, e.g.][]{binney_tremaine}
\begin{equation}
	\label{eq:aniso}
	1 - \frac{\disp[tan]^2}{2 \disp[rad]^2} = \frac{(r/\ra)^2}{1 + (r/\ra)^2} \,,
\end{equation}
which we set up numerically with the tool \textsc{agama} \citep{agama}.
These models are approximately isotropic in the inner regions (\hbox{$r<\ra$}) and become increasingly radially anisotropic as the distance from the centre of the cluster increases beyond the anisotropy radius, $\ra$.
Thus the model which is initially isotropic has $\ra{\rightarrow}\infty$, while for the radially anisotropic ones, we set $\ra = \rh[,0]$.
See Table~\ref{tab:models} for the summary of all model parameters and the naming convention adopted here.

\section{Results}
\label{sec:results}

First, we focus on the evolution towards core collapse and EEP in models with different velocity distributions, different filling factors, and no primordial binaries. New models of this study are also compared with the results from \citetalias{pav_ves_letter} (i.e., models \texttt{i0.1s} and \texttt{a0.1s}). We then continue with the models that include a population of primordial binary stars. Finally, we present the results of the analysis concerning the time evolution of the various EEP parameters calculated by means of the projected quantities measured analogously to the observational studies.

\subsection{Core collapse}

\begin{figure*}
	\centering
	\includegraphics[width=.495\linewidth]{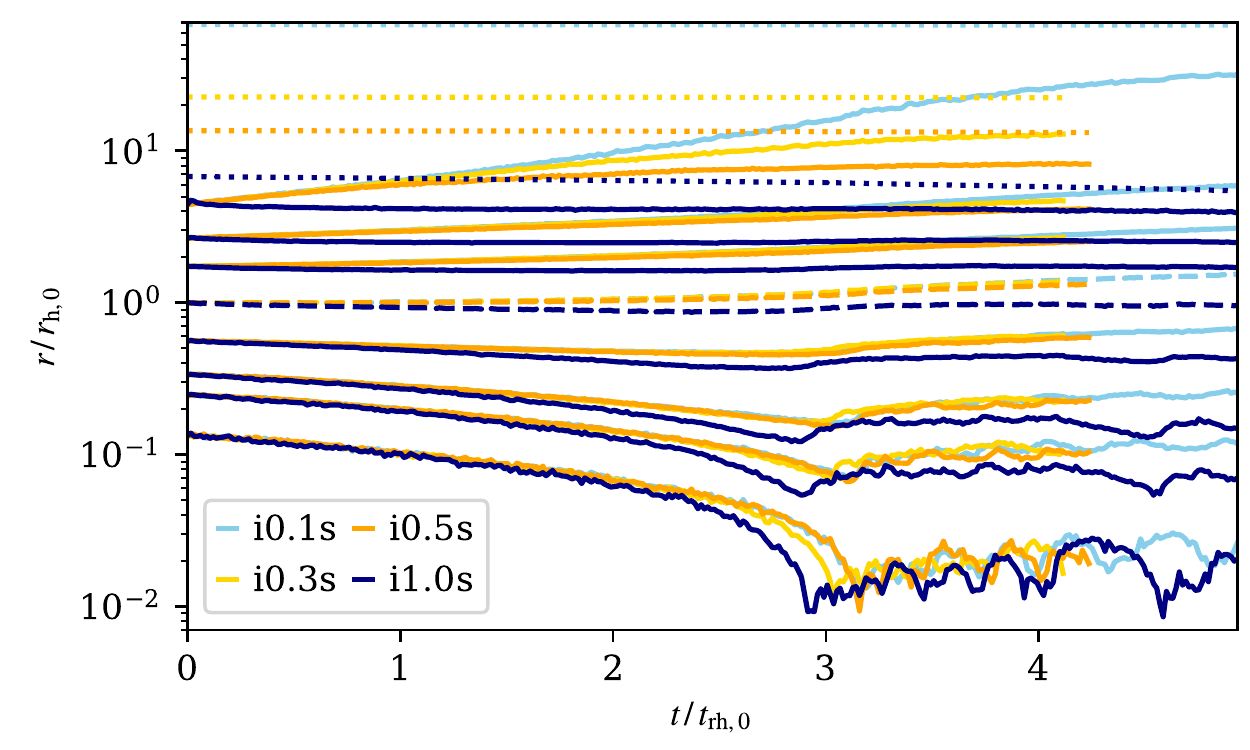}\hfill
	\includegraphics[width=.495\linewidth]{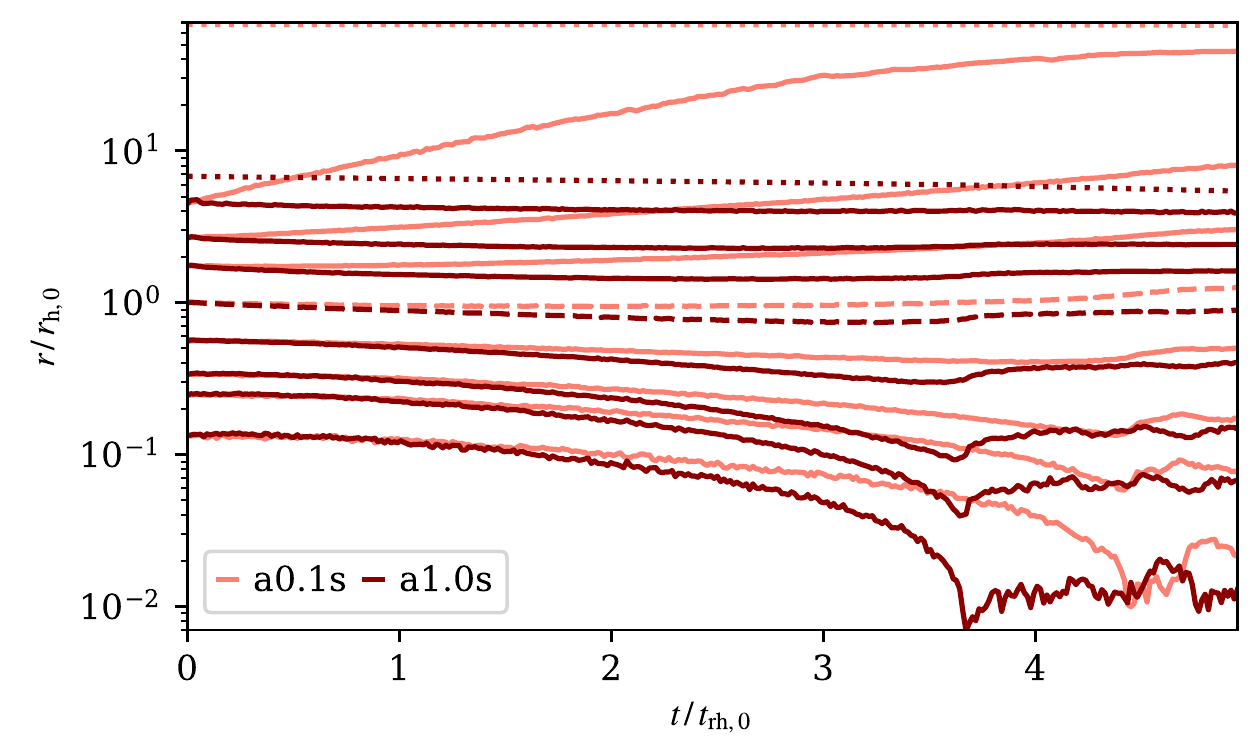}
	
	\caption{Time evolution of the isotropic (\textbf{left}) and anisotropic (\textbf{right}) models without binaries shown by the 1, 5, 10, 25, 50, 75, 90 and 99\,\% Lagrangian radii (normalised to the initial half-mass radius -- plotted with a dashed line). The tidal radius of each model is plotted with a dotted line. Time is in units of the initial half-mass relaxation time.}
	\label{fig:lagr}
	
	\vspace{\floatsep}
	
	\includegraphics[width=.495\linewidth]{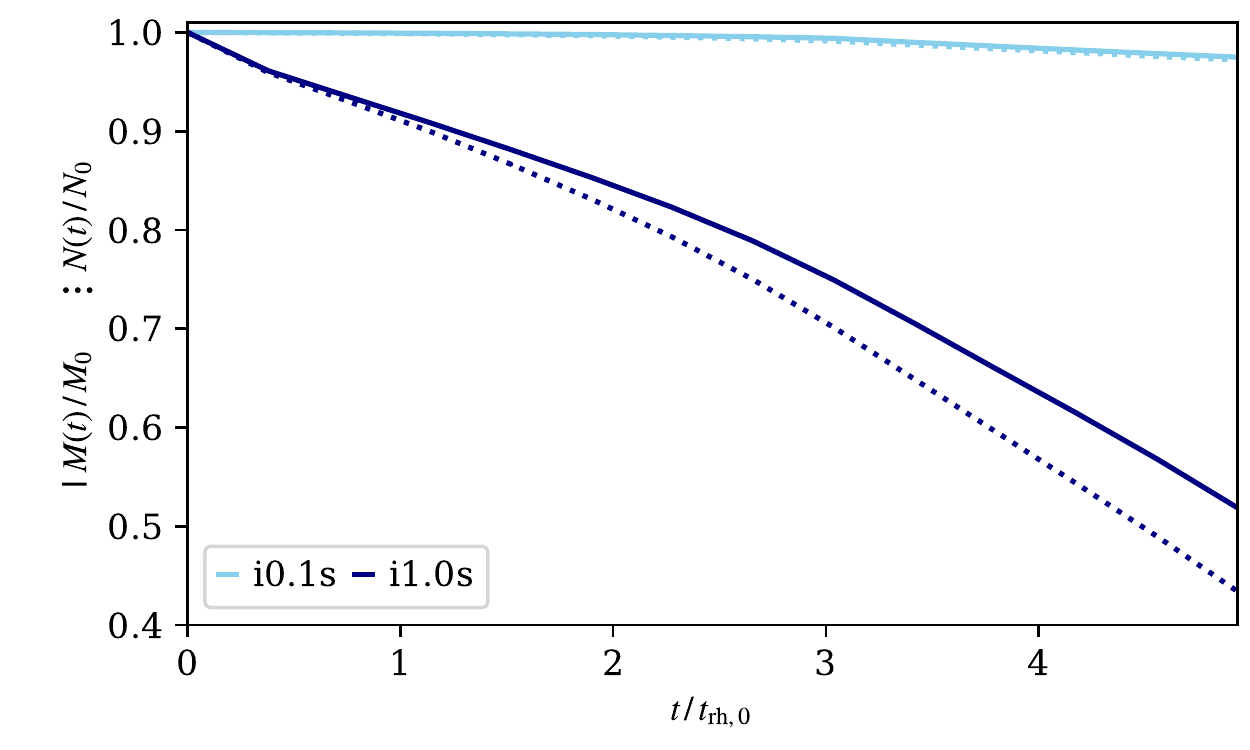}\hfill
	\includegraphics[width=.495\linewidth]{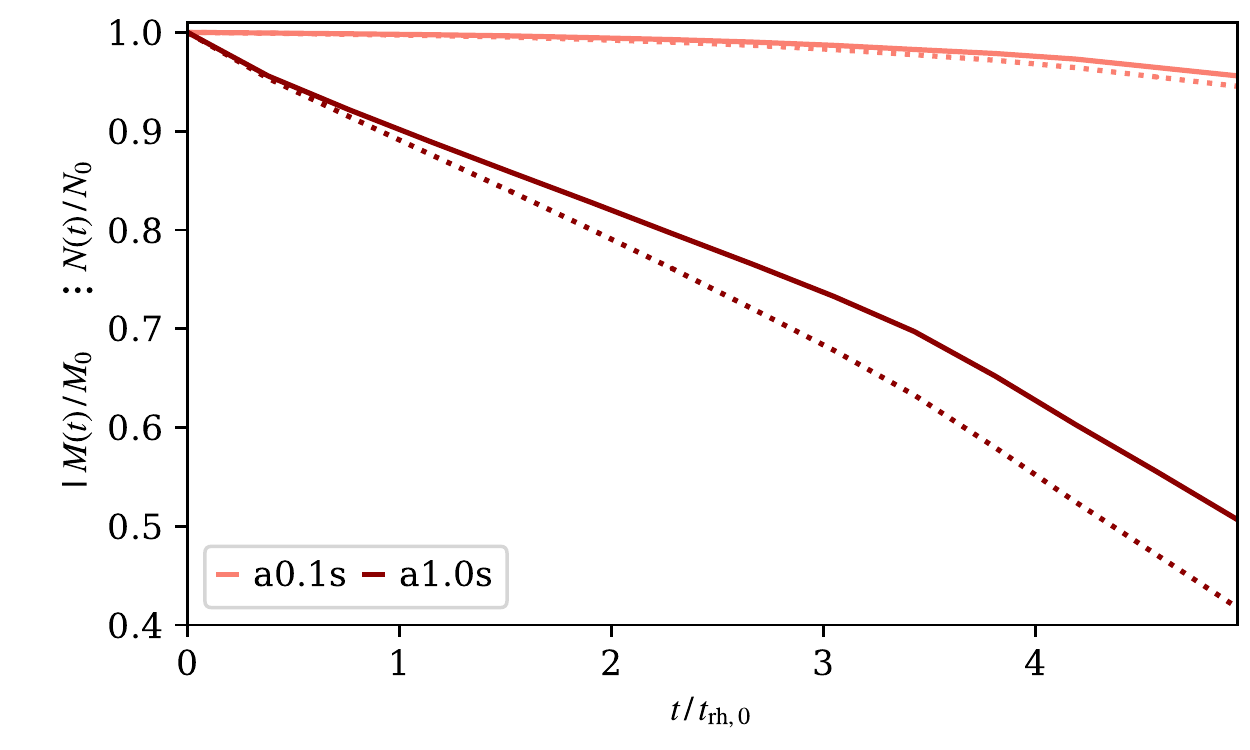}
	
	\caption{Time evolution of the total mass (solid lines) and number of stars (dotted lines) in the modelled clusters -- isotropic (\textbf{left}) and anisotropic (\textbf{right}). Both quantities are expressed as fractions of the initial $M_0$ and $N_0$, respectively. Only the extreme filling factors of 0.1 and 1.0 are shown.}
	\label{fig:mass}
\end{figure*}

In \citetalias{pav_ves_letter}, we showed that the velocity anisotropy slows down the evolution towards core collapse in the tidally underfilling models \citep[a similar result has also been discussed by][]{breen_var_heg}. Here, we extend that analysis and consider different strengths of the external tidal field. In Fig.~\ref{fig:lagr}, we plot the time evolution of various Lagrangian radii for the tidally underfilling and filling models; for the isotropic initial conditions we also show the evolution of two intermediate filling factors.
A comparison between the evolution of the \texttt{i1.0s} and the \texttt{a1.0s} models provides evidence that a trend similar to that reported in \citetalias{pav_ves_letter} is present also for the tidally filling models. Fig.~\ref{fig:lagr} also shows that all the initially isotropic models reach core collapse at about the same time (in units of $\trh[,0]$), with the tidally filling model collapsing slightly earlier than the others \citep[the time of core collapse, $\tcc$, was determined following the method of][]{pavl_subr}. So, for the set of initial conditions considered here, the time scale for the core evolution in isotropic systems is almost independent of the tidal field. This is in agreement with the results of \citet{giersz_heggie96,giersz_heggie97} who found similar core collapse times for isolated and tidally limited models, using different initial density profiles and mass function from those in our study.
In the initially anisotropic models, on the other hand, core collapse is reached significantly earlier in the tidally filling model -- a gap of about $1\,\trh[,0]$. Although the choice of the initial relaxation time as a time unit does not take into account differences in the mass-loss rate of the models (and therefore does not fully reflect the relaxation process itself), in our case, mass loss alone cannot explain the various trends in the times of core collapse for isotropic and anisotropic models completely. As shown in Fig.~\ref{fig:mass}, both the isotropic (left-hand panel) and the anisotropic models (right-hand panel) show a significant difference between the mass loss of the filling and the underfilling models but only the anisotropic models exhibits considerable differences in $\tcc$. We will explore the possible role of anisotropy on this phenomenon in our upcoming work (Pavlík et al. in prep).

\subsection{Energy equipartition}

\begin{figure*}
	\centering
	
	\includegraphics[width=\linewidth]{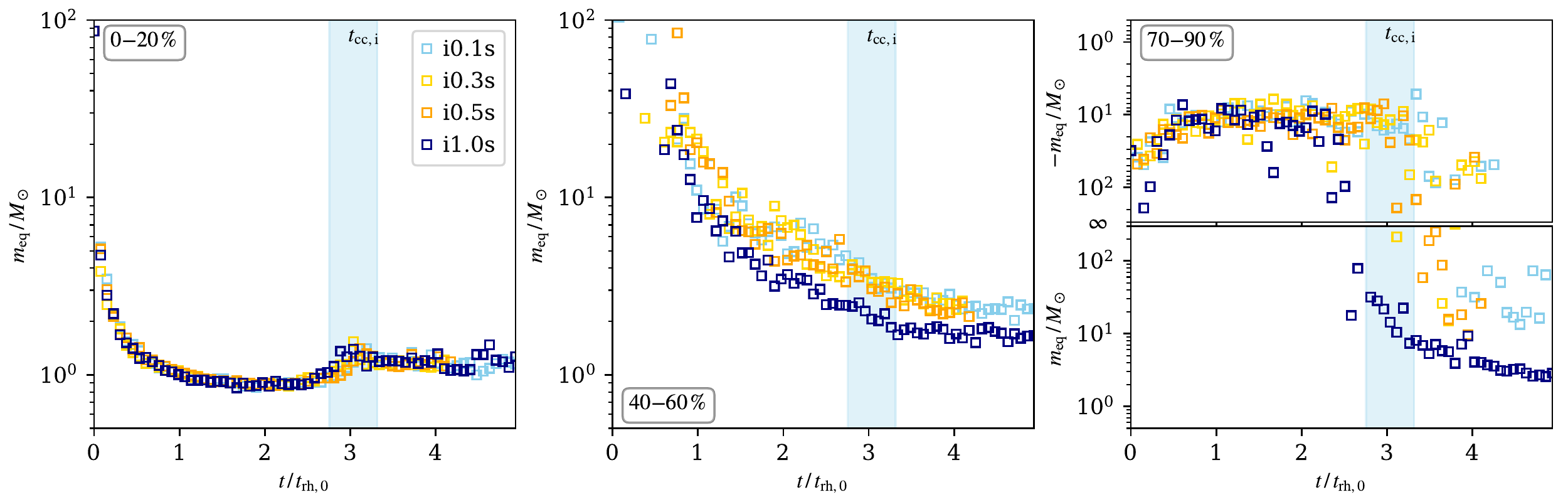}
	
	\caption{Time evolution of the equipartition mass in the isotropic models with different filling factors, see Table~\ref{tab:models} for the notation.  The approximate time of core collapse of all the models is shown by the shaded area. Each panel (from left to right) represents the Lagrangian shell of a given percentage determined separately for each model. The left-hand and centre plots show the evolution of $\meq$ in the positive values only; the right-hand side plot is split vertically by a double line which represents the limit $\meq{\rightarrow}\pm\infty$ that appears between the positive (lower plot) and negative (upper plot) values of $\meq$. The design of Figs.~\ref{fig:meq_iso_fix}, \ref{fig:meq_aniso}, \ref{fig:meq_bin} \&~\ref{fig:meq_proj} is similar (and adopted from \citetalias{pav_ves_letter}).}
	\label{fig:meq_iso}
	
	\vspace{\floatsep}

	\includegraphics[width=\linewidth]{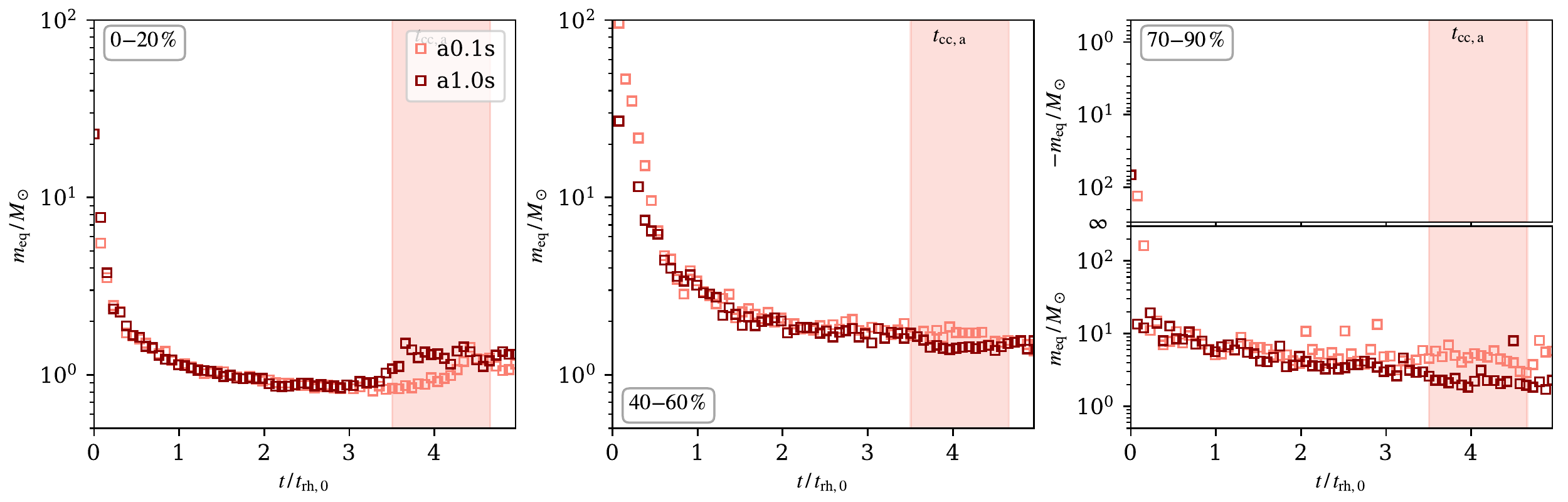}
	
	\caption{Same as Fig.~\ref{fig:meq_iso} but for the initially anisotropic models.}
	\label{fig:meq_aniso}
	
	\vspace{\floatsep}
	
	\includegraphics[width=\linewidth]{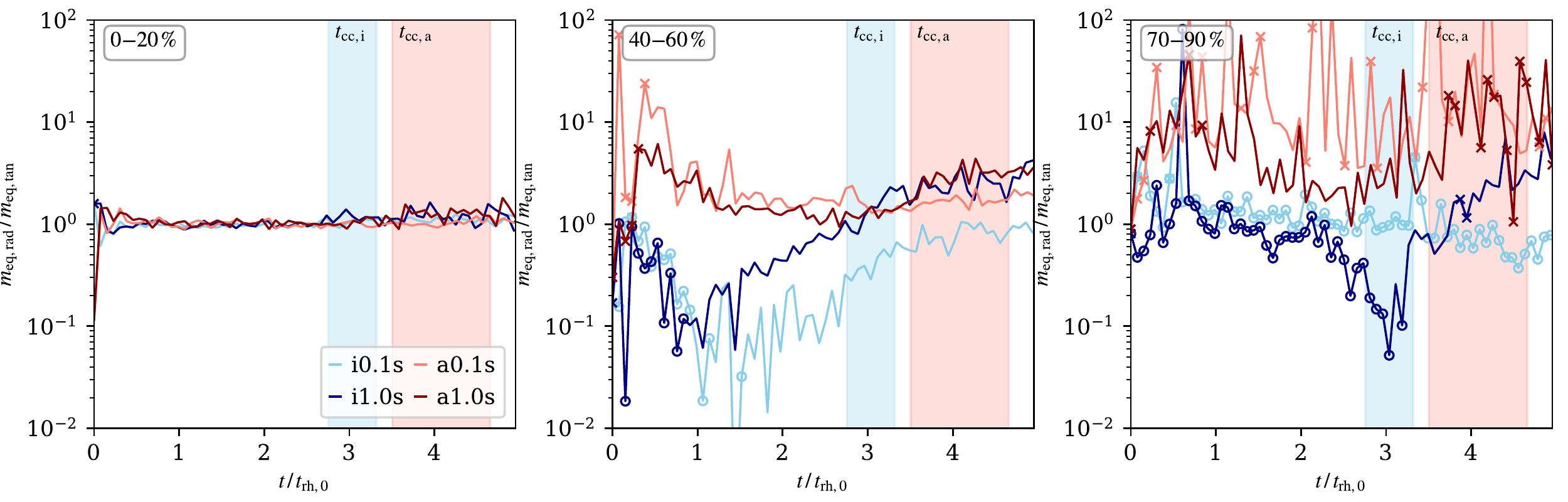}
	
	\caption{Time evolution of the ratio between the radial and tangential components of the equipartition mass in the models with single stars. Each line represents the absolute value of the ratio $\meq[,rad]/\meq[,tan]$. To distinguish the sign of the numerator and the denominator, we use the following symbols: the times when $\meq[,rad]{<}0$ are marked by crosses (\boldmath$\times$\unboldmath), $\meq[,tan]{<}0$ by circles (\boldmath$\bigcirc$\unboldmath), and if both values are positive, only a line with no symbol is plotted. The initially isotropic and anisotropic models are distinguished by colour and the filling factors by the colour opacity, see also Table~\ref{tab:models} for the notation. The approximate time of core collapse of the models is shown by the shaded area.}
	\label{fig:meq_ratio}
\end{figure*}

\begin{figure*}
	\centering
	\includegraphics[width=.96\linewidth]{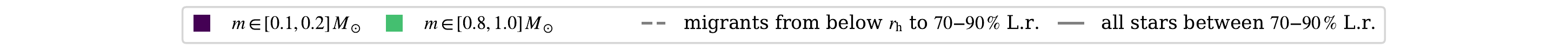}\\
	
	\includegraphics[width=.249\linewidth]{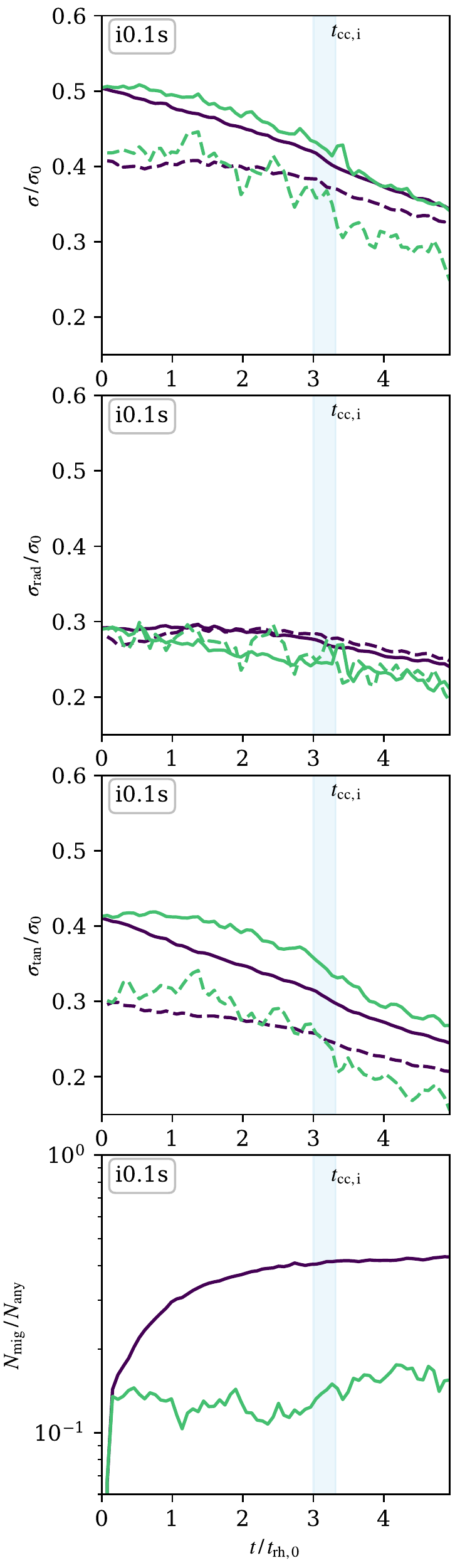}\hfill
	\includegraphics[width=.249\linewidth]{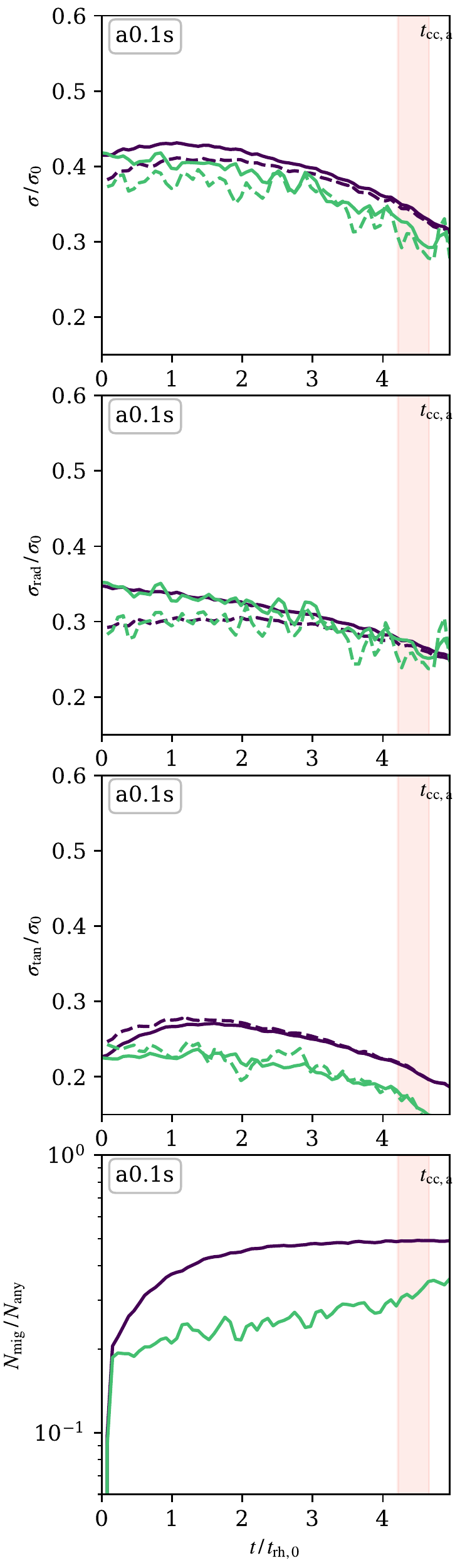}\hfill
	\includegraphics[width=.249\linewidth]{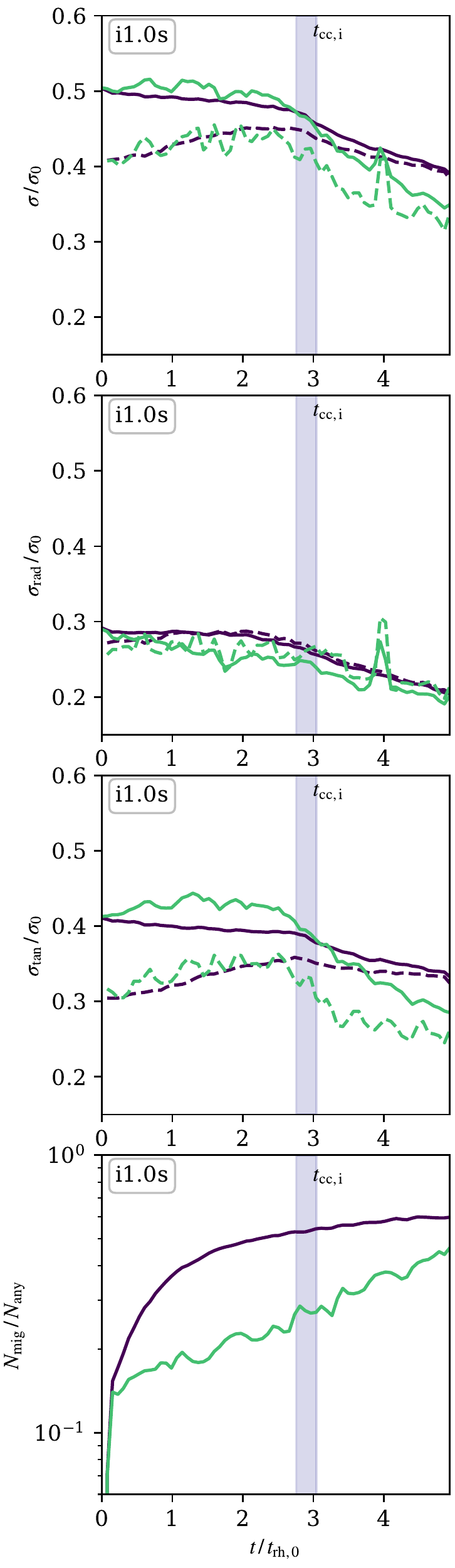}\hfill
	\includegraphics[width=.249\linewidth]{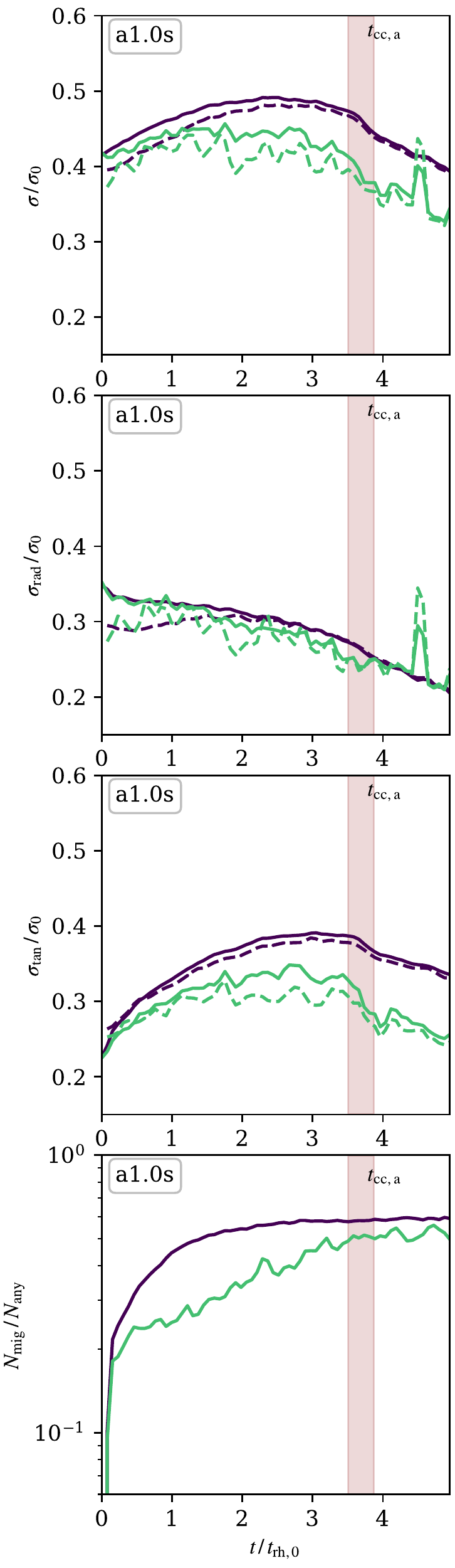}
	\caption{\textbf{Top three rows:} Total velocity dispersion, $\disp = \sqrt{\disp[rad]^2 + \disp[tan]^2}$, and its radial ($\disp[rad]$) and tangential ($\disp[tan]$) components in the outer regions (i.e., between $70$ and $90\,\%$ Lagrangian radii) of the models without primordial binaries. Each model is plotted in one column and labelled according to Table~\ref{tab:models}. Two extreme mass groups are plotted in each panel: $m/\Msun \leq 0.2$ and $\geq 0.8$ (distinguished by line colours). Two populations of stars are also compared in each mass group -- all stars that are in the $70{-}90\,\%$ region at time $t$ (solid lines) and those stars that are in this region at time $t$ but were initially in the region below the half-mass radius (dashed lines).
	\textbf{Bottom row:} The ratio between the migrated population and stars with any origin; also colour coded by mass. The approximate time of core collapse of each model is shown in each panel by the vertical colour shaded strip.}
	\label{fig:mig}
\end{figure*}

\begin{figure*}
	\centering
	
	\includegraphics[width=\linewidth]{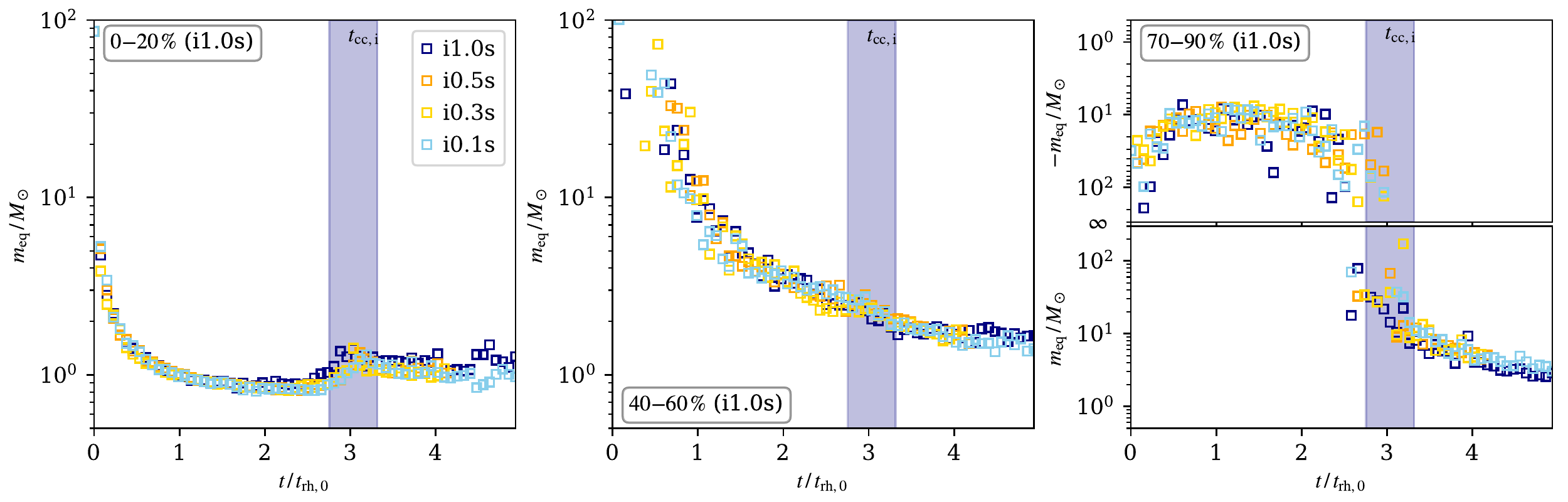}
	
	\caption{Same as Fig.~\ref{fig:meq_iso} but each Lagrangian shell is determined from the \texttt{i1.0s} model and then kept fixed in the other models.}
	\label{fig:meq_iso_fix}
\end{figure*}

In order to quantify the evolution towards EEP, we adopt a kinematic parameter $\meq$ \citep[the equipartition mass; introduced by][]{bianchini_meq} which is defined as
\begin{equation}
	\label{eq:meq}
	\disp(r,m) \propto \exp{[-m / (2\meq)]} \,,
\end{equation}
where $\disp$ is the velocity dispersion for different radial and mass bins.
Fig.~\ref{fig:meq_iso}, shows that in the case of the isotropic models, the time evolution of $\meq$ is qualitatively similar in star clusters with different filling factors. The same is also true for the evolution of $\meq$ in the tidally filling and underfilling models that started with a radially anisotropic velocity distribution (see Fig.~\ref{fig:meq_aniso}).

As for the initially isotropic clusters, similarly to the underfilling model presented in \citetalias{pav_ves_letter} (cf. Fig.~3 in there), all of them start with high $\meq$ (ideally infinite). In the core region, the system evolves towards EEP and $\meq$ decreases (see the left-hand panel of Fig.~\ref{fig:meq_iso} here), while the cluster halo evolves away from EEP as indicated by negative $\meq$ (see the right-hand panel of the same figure).
In Fig.~\ref{fig:meq_ratio}, we further explore the evolution towards EEP and its dependence on the velocity dispersion components by showing the time evolution of the ratio of $\meq[,rad]$ (calculated using the radial velocity dispersion) and $\meq[,tan]$ (using the tangential one). As shown in this figure, for both isotropic and anisotropic systems the evolution towards EEP in the intermediate and outer regions does depend on the component of the velocity dispersion.

The inverted trend observed in the outer regions of the isotropic systems is another manifestation of the dependence of the evolution towards EEP on the component of the velocity dispersion. In particular, this inversion is caused by its tangential component. To show that, we plot the total velocity dispersion and its components, $\disp[rad]$ and $\disp[tan]$, in the outer cluster regions separately for two mass groups in the top three rows of Fig.~\ref{fig:mig}. In the isotropic models -- \texttt{i0.1s} (first column) and \texttt{i1.0s} (third column) -- $\disp[tan]$ of the higher-mass stars (green solid line) is larger than $\disp[tan]$ of the lower-mass stars (purple solid line), which is the opposite of the standard evolution towards EEP found, for example, in the anisotropic models (second and fourth columns). The total $\disp$ of the higher-mass stars in the initially isotropic models is also larger than $\disp$ of the lower-mass population (see the top panels of Fig.~\ref{fig:mig}).
Fig.~\ref{fig:meq_ratio} further illustrates how these components of the velocity dispersion affect the value of $\meq$: only the blue lines, which represent the initially isotropic models, have a systematically negative tangential component (marked by circles) in the outer regions (right-hand panel).\footnote{We note, that there is no simple conversion between the total $\meq$ and its $\meq[,rad]$ or $\meq[,tan]$ components, and thus, also not between the data points plotted in Fig.~\ref{fig:meq_ratio} and Figs.~\ref{fig:meq_iso} \&~\ref{fig:meq_aniso}.}

We interpret this in the following way. Due to two-body relaxation, low-mass stars preferentially migrate from the inner to the outer regions on highly eccentric/radial orbits leading to the development of a radially anisotropic velocity distribution \citep[see, e.g.,][]{giersz_heggie96,Tio_Ves_Var16}. As this process continues, the increasing fraction of migrant low-mass stars in the outer population becomes responsible for the observed more rapid decline of the tangential velocity dispersion which leads to inverted EEP observed in our simulations. As more higher-mass stars continue to migrate in the outer regions over time, especially around core collapse, the differences in $\disp[tan]$ between low-mass and high-mass stars gradually decrease and the cluster starts to evolve towards the expected trend between $\disp$ and $m$.

This process acts more efficiently in the isotropic and underfilling systems. For the isotropic filling system, the higher mass-loss rate, especially of stars on radial/highly-eccentric orbits leads to a shorter inverted phase (compare the evolution of $\meq$ for the \texttt{i0.1s} and \texttt{i1.0s} models in the right panel of Fig.~\ref{fig:meq_iso}).

To further illustrate the role of mass loss and the implications of different filling factors, we show the evolution of $\meq$ for various initially isotropic systems calculated all at the same radial distances instead of adopting the Lagrangian radii defined separately for each model. This allows for a more consistent comparison of the internal evolution. For that we chose the Lagrangian radii of the filling model, \texttt{i1.0s} (but any set of radial shells fixed for all models could be chosen).
We note that all the clusters  we studied have the same initial spatial distribution but while the filling models cannot evolve beyond their initial truncation radius (which coincides with their tidal radius), the underfilling models can expand up to ten times larger radii their initial size (see Fig.~\ref{fig:lagr}). So by comparing the values of $\meq$ at the same clustercentric distances determined by the model \texttt{i1.0s}, the outer regions of the less-filling clusters are now excluded.
In all models, the evolution of $\meq$ is essentially identical in the overlapping regions, regardless of the initial filling factors (see Fig.~\ref{fig:meq_iso_fix}).
Hence, the main difference in the evolution towards (or away from) EEP between the tidally filling and underfilling models is when we compare the outer regions of the filling clusters with the outer regions of the underfilling ones. The latter are populated by stars expanding beyond the cluster's initial truncation radius and can keep a diffuse population of preferentially low-mass stars on very eccentric/radial orbits which forms a radially anisotropic halo. On the other hand, the outer regions of the tidally filling cluster are limited by the initial truncation radius and contain the native isotropic stellar population mixed with the escaping stars, therefore, in this case, such an anisotropic halo never forms.

\begin{figure*}
	\centering
	
	\includegraphics[width=.495\linewidth]{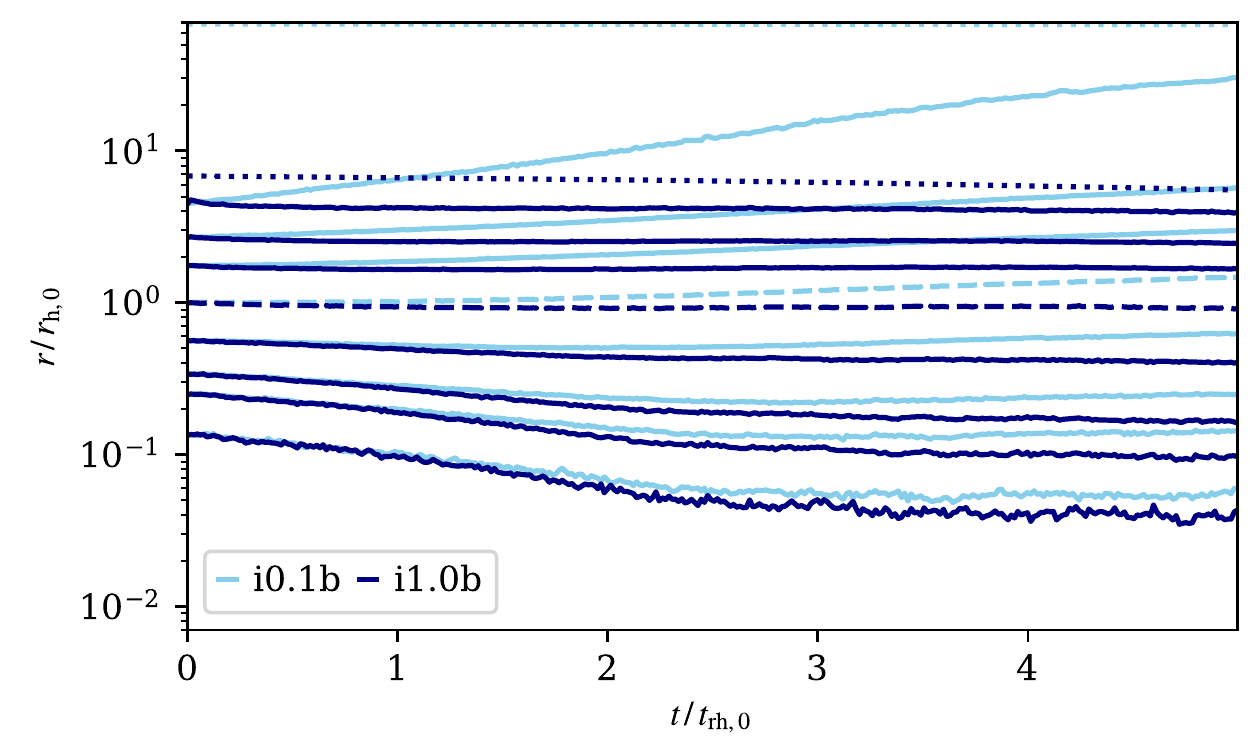}\hfill
	\includegraphics[width=.495\linewidth]{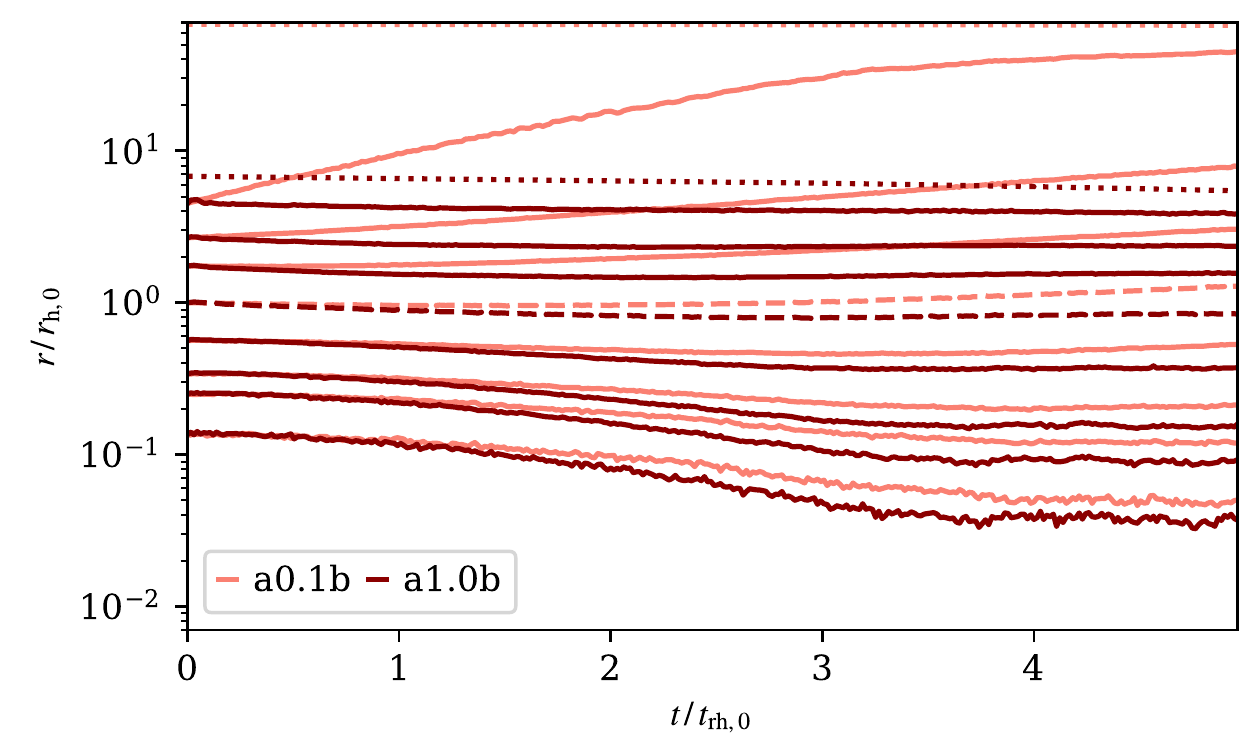}
	
	\caption{Same as Fig.~\ref{fig:lagr} but for the models with primordial binaries.}
	\label{fig:lagr_bin}
\end{figure*}
\begin{figure*}
	\centering
	
	\includegraphics[width=\linewidth,trim=0 18 0 1,clip]{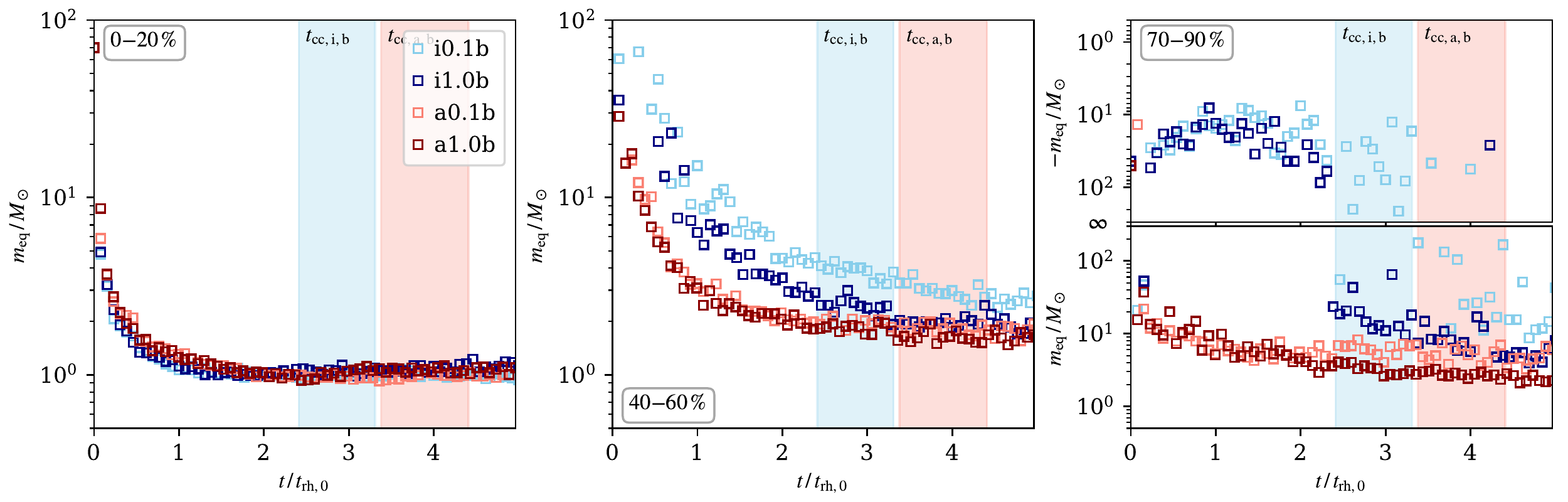}\\
	\includegraphics[width=\linewidth,trim=0 5 0 1,clip]{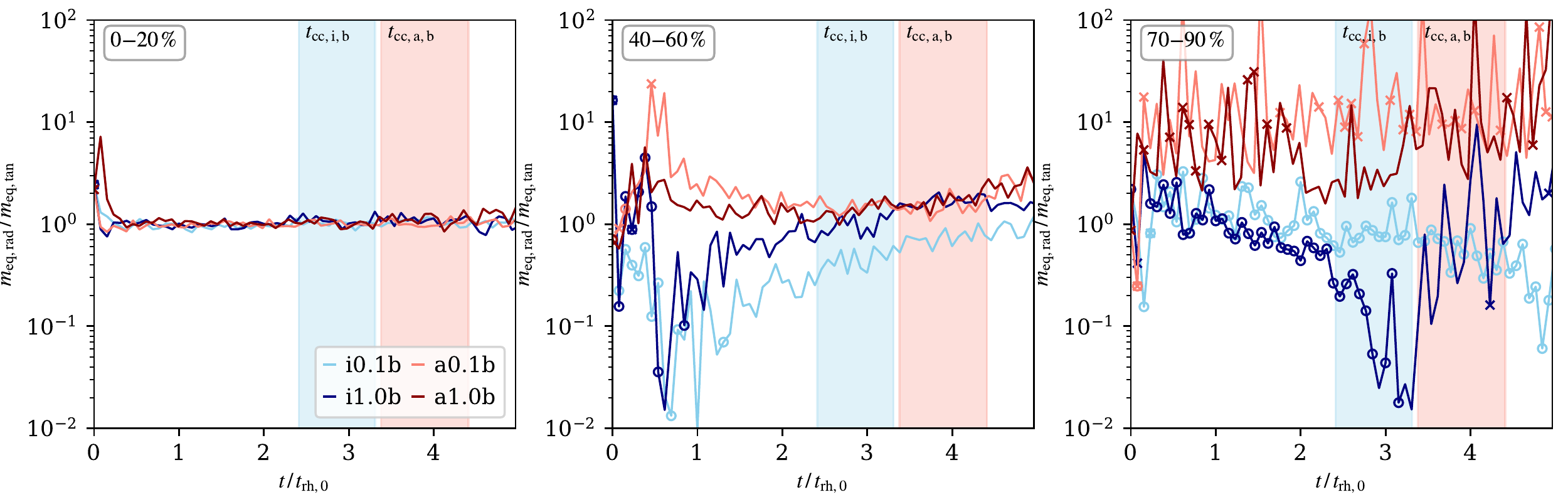}
	
	\caption{\textbf{Top:} Time evolution of the equipartition mass in the models with binaries. The initially isotropic and anisotropic models are distinguished by colour and the filling factors by the colour opacity. See also Table~\ref{tab:models} for the notation and Fig.~\ref{fig:meq_iso} for the explanation of the figure layout. The approximate time of core collapse of all the models is shown by the shaded area.
	\textbf{Bottom:} Same as Fig.~\ref{fig:meq_ratio} but for the models with primordial binaries.}
	\label{fig:meq_bin}
\end{figure*}

In the initially anisotropic systems, on the other hand, the outer regions are dominated by stars on eccentric radial orbits. Therefore, the effect of the migrating low-mass population on the outer regions is smaller and the systems never evolve towards inverted EEP (see the right-hand panel of Fig.~\ref{fig:meq_aniso}). The main difference between the initially isotropic and anisotropic models is that the change of velocity dispersion in the outer shells of the former has mainly a dynamical origin due to the diffusion of preferentially low-mass stars, whereas the latter already started with a population of stars that are, independently of their masses, on eccentric radial orbits. Although there is some dynamical diffusion of stars in the anisotropic model as well, it has a considerably lower effect on the change of $\disp[rad]$ or $\disp[tan]$ for stars with different masses. This can be seen in the bottom row of Fig.~\ref{fig:mig} where we plot the ratio of stars that migrated to the outer regions of the cluster and the total number of stars that are in the region at a given time. When comparing the two underfilling models (two bottom-left panels), or the two filling models (two bottom right panels) within the first fraction of a relaxation time, the anisotropic models generally show a more rapid outward migration of all stars associated with the initial anisotropic velocity distribution.
Finally, we note that, similarly to what was found in \citetalias{pav_ves_letter} for the underfilling models, even the evolution of the filling models towards EEP is more rapid in the initially anisotropic models than the isotropic ones.

\subsection{Binary stars}

\begin{table}
	\centering
	\caption{The minimum values of $\meq$ in the inner Lagrangian shell (0--20\%) of our models with and without primodial binaries.}
	\begin{tabular}{lcp{1em}lc}
		\hline
		model & $\min{\meq}$ && model & $\min{\meq}$ \\
		\hline
		\texttt{i0.1s} & $0.85$ &&\texttt{i0.1b} & $0.95$ \\
		\texttt{i1.0s} & $0.85$ &&\texttt{i1.0b} & $0.99$ \\
		\texttt{a0.1s} & $0.83$ &&\texttt{a0.1b} & $0.93$ \\
		\texttt{a1.0s} & $0.86$ &&\texttt{a1.0b} & $0.94$ \\
		\hline
	\end{tabular}
	\label{tab:min_meq}
\end{table}

In this section, we focus on the models with primordial binary stars -- we have set up four different combinations of the initial conditions (isotropic and anisotropic; filling factors 0.1 and 1.0, see Table~\ref{tab:models}). Primordial binaries halt core collapse earlier and the core does not contract as much as in the models where binaries had to form dynamically \citep[compare Fig.~\ref{fig:lagr} and Fig.~\ref{fig:lagr_bin}; see also, e.g.,][] {tre_heg_hut}. Hence, the precise identification of the time when the core contraction stops is more difficult and we only determined $\tcc$ approximately from the minimum of the 1\,\% Lagrangian radius.

The early termination of the core contraction due to the presence of primordial binaries also has implications on the evolution towards EEP. Clusters with a primordial population of binaries do not reach values of $\meq$ as low as clusters without them, therefore, the former do not reach a dynamical state as close to equipartition as the latter (compare the top left panel of Fig.~\ref{fig:meq_bin} and the left-hand panels of Figs.~\ref{fig:meq_iso} \&~\ref{fig:meq_aniso}; see also Table~\ref{tab:min_meq})

In contrast, the overall trends persist and we reach the same conclusions we discussed for the models with only single stars. Namely, the evolution towards EEP depends on the selected velocity dispersion component in the intermediate and outer regions (see the bottom middle and right-hand panels of Fig.~\ref{fig:meq_bin}); the isotropic models with binaries (\texttt{i0.1b} and \texttt{i1.0b}) evolve towards a state of inverted EEP\footnote{We note that the isotropic models with binaries also do not reach positive values as low as those in the isotropic models with only single stars.} but the anisotropic models with binaries (\texttt{a0.1b} and \texttt{a1.0b}) show no such trend (compare the right-hand panels of Figs.~\ref{fig:meq_iso} \&~\ref{fig:meq_aniso} and the top panel of Fig.~\ref{fig:meq_bin}). As in the models with single stars, it is the tangential component of the velocity dispersion and, hence, the tangential component of the equipartition mass that is responsible for this behaviour (see the bottom right panel of Fig.~\ref{fig:meq_ratio}) since it has systematically negative values (marked by circles). We note that the radial component of the initially anisotropic models also reaches negative values but these are just due to occasional fluctuations. We also notice that the change of sign of $\meq$ in the isotropic models (and the bifurcation of the anisotropic models) happens again at approximately the time of core collapse (see again the top row of Fig.~\ref{fig:meq_bin}). Similarly to what found for the single stars models, even here, the evolution of $\meq$ in the intermediate and outer regions is more rapid in the initially anisotropic systems than the isotropic ones.

\subsection{Line-of-sight projection}

\begin{figure*}
	\centering
	
	\includegraphics[width=\linewidth,trim=0 18 0 1,clip]{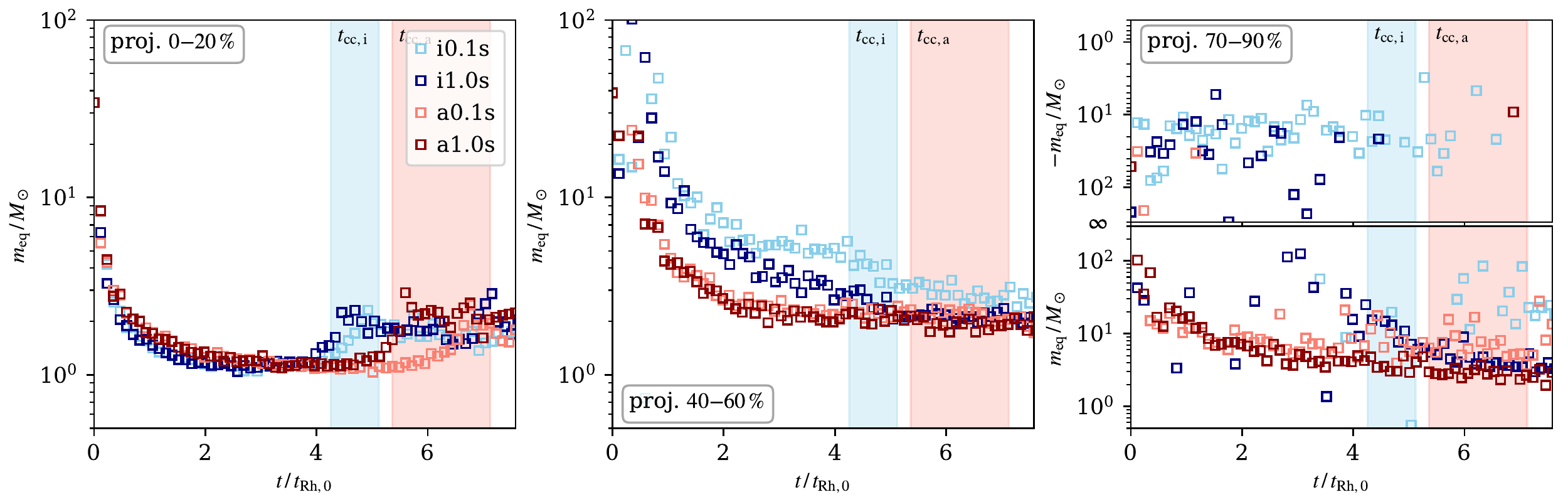}\\
	\includegraphics[width=\linewidth,trim=0 5 0 1,clip]{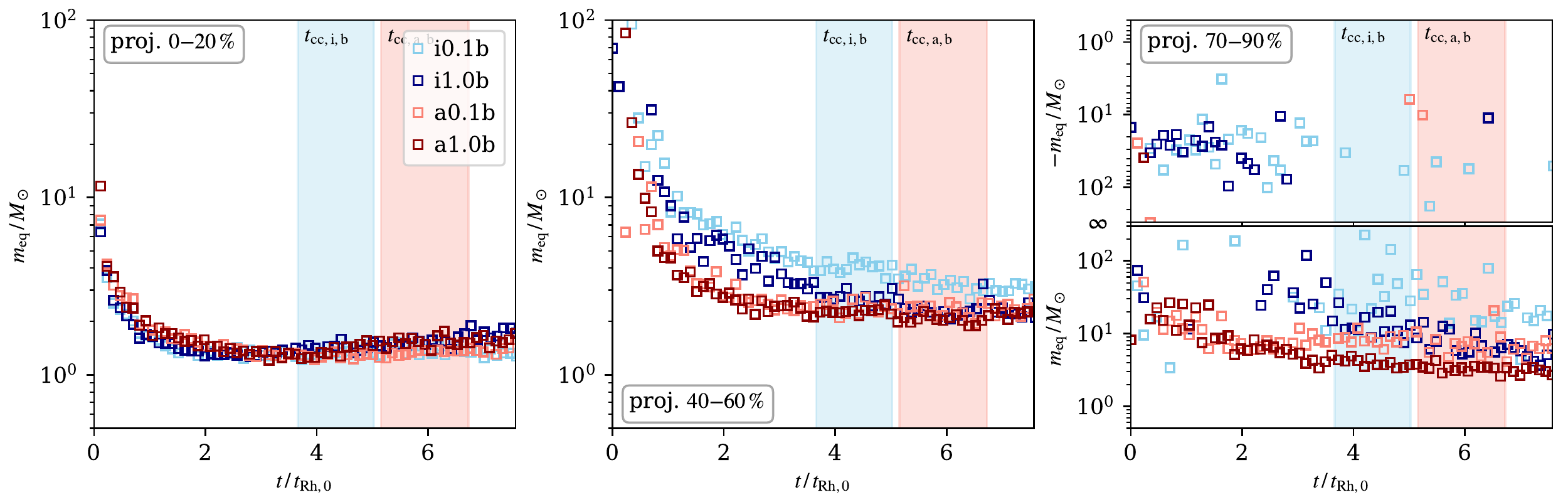}
	
	\caption{Time evolution of the equipartition mass in the models with (\textbf{bottom}) and without binaries (\textbf{top}), viewed in projection. We plot the average $\meq$ calculated from the values on the $xy$, $xz$ and $yz$ planes. See Table~\ref{tab:models} for the notation of the models and Fig.~\ref{fig:meq_iso} for the explanation of the figure layout. Time is expressed in the initial half-mass relaxation times calculated from the projected initial half-mass radius.}
	\label{fig:meq_proj}
	
	\vspace{0.1\floatsep}
	
	\includegraphics[width=\linewidth,trim=0 18 0 1,clip]{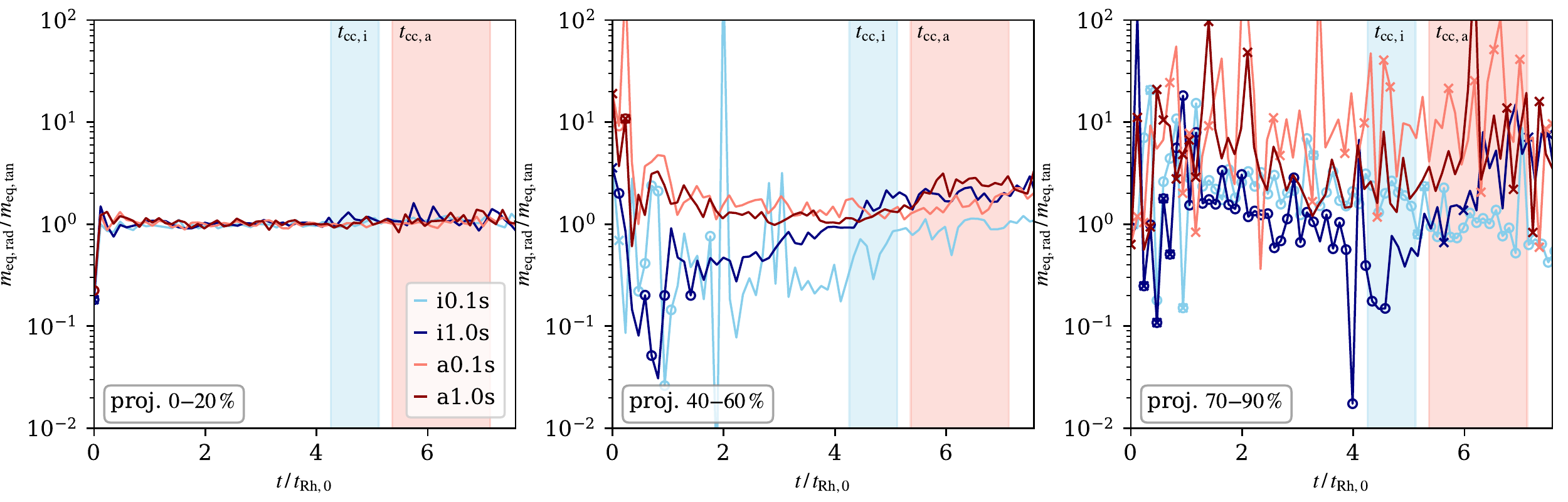}\\
	\includegraphics[width=\linewidth,trim=0 5 0 1,clip]{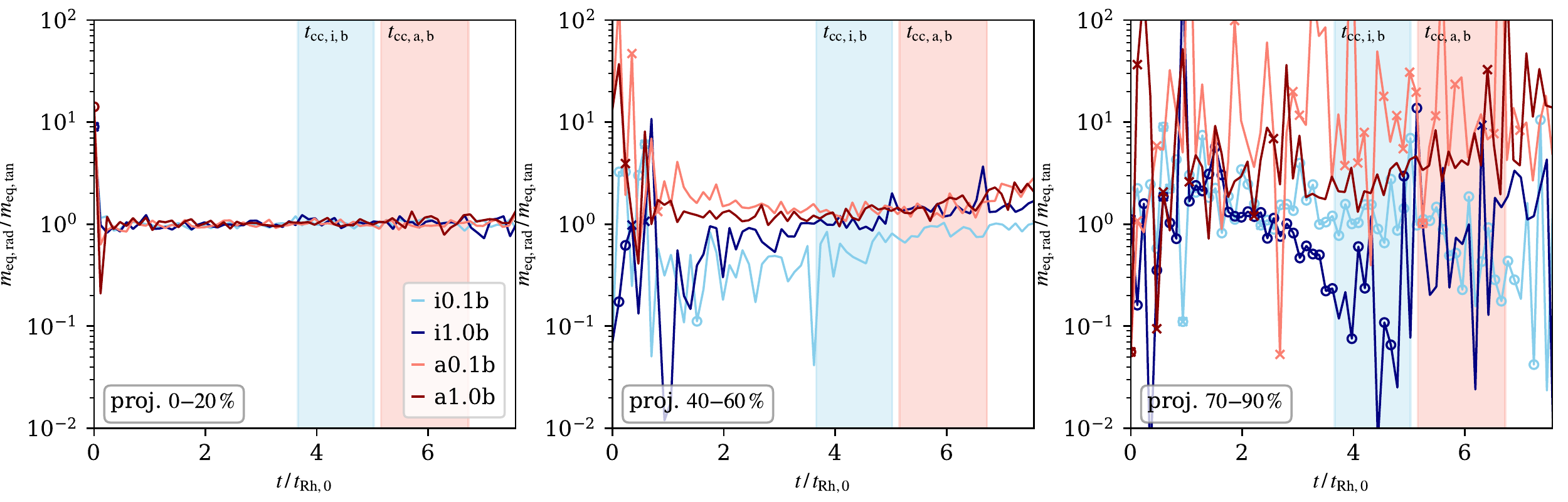}
	
	\caption{Same as Fig.~\ref{fig:meq_ratio} but for the projected radial and tangential components of $\meq$ (see also Fig.~\ref{fig:meq_proj}).}
	\label{fig:meq_proj_ratio}
\end{figure*}

\begin{figure*}
	\centering
	
	\includegraphics[width=\linewidth,trim=0 18 0 1,clip]{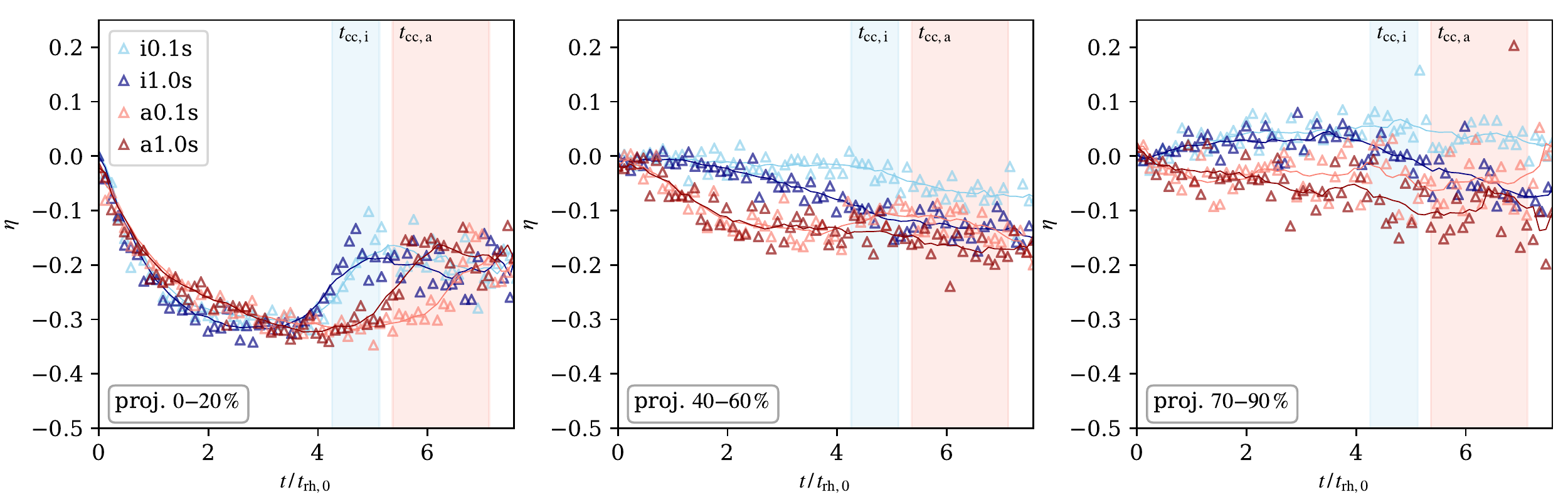}\\
	\includegraphics[width=\linewidth,trim=0 5 0 1,clip]{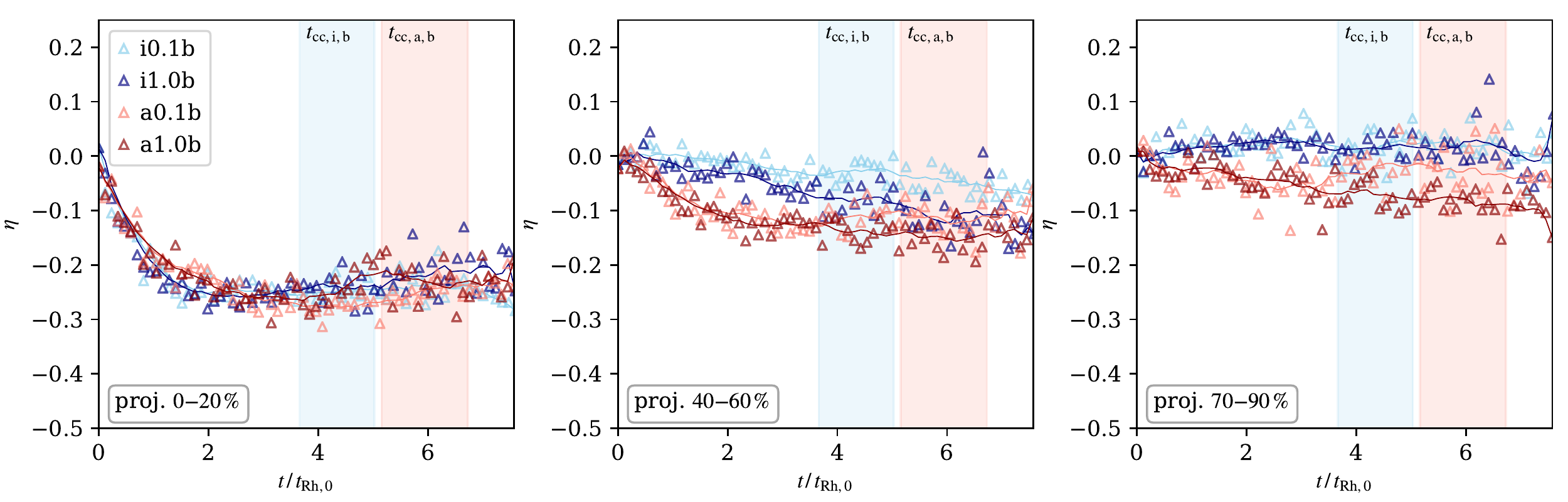}
	
	\caption{Same as Fig.~\ref{fig:meq_proj} but for the parameter $\etam$. This plot was made using stars of masses $0.4 \leq m/\Msun \leq 1.0$. The thin lines show the moving averages and are added to guide the eye.}
	\label{fig:eta_proj}

	\vspace{0.1\floatsep}
	
	\includegraphics[width=\linewidth,trim=0 18 0 1,clip]{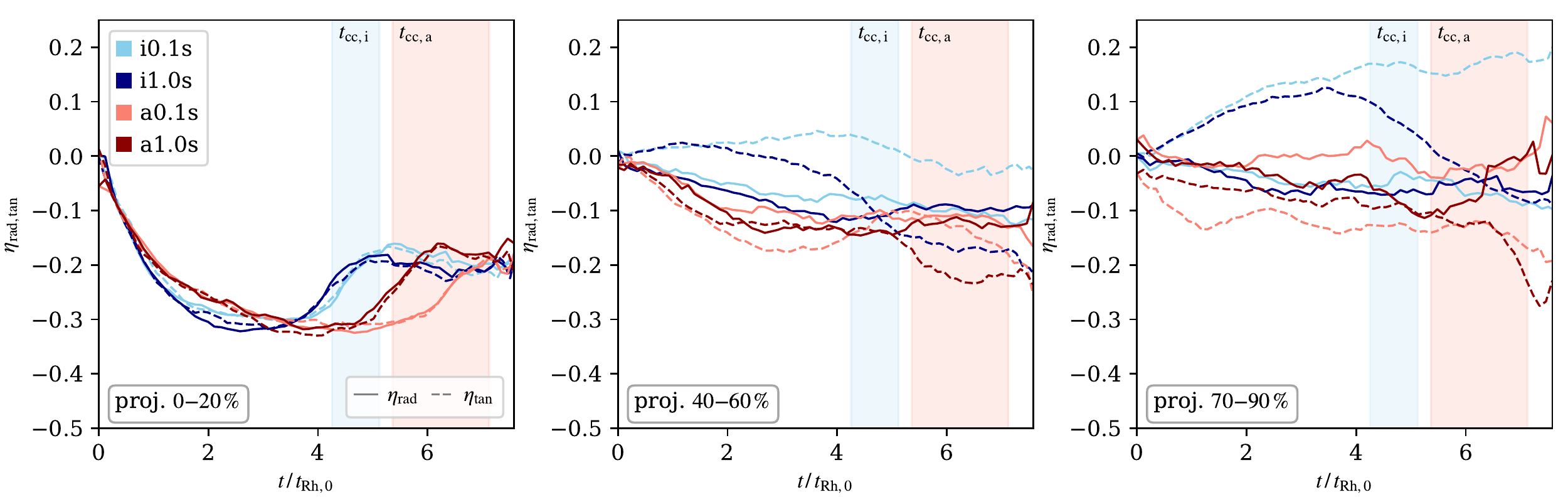}\\
	\includegraphics[width=\linewidth,trim=0 5 0 1,clip]{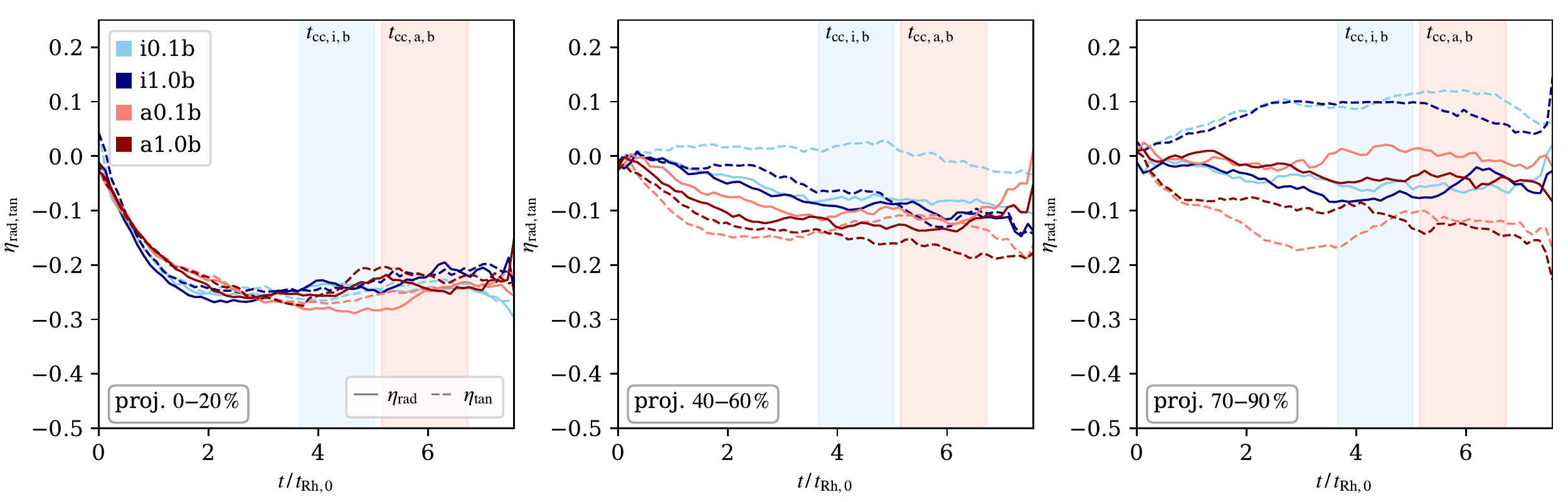}
	
	\caption{Same as Fig.~\ref{fig:eta_proj} but for the radial and tangential components of $\etam$. We note that the oscillations between positive and negative values (as seen in Fig.~\ref{fig:eta_proj} or with $\meq$ in Fig.~\ref{fig:meq_proj}) are not visible here since only the moving averages are shown.}
	\label{fig:eta_proj_radtan}
\end{figure*}

Until now, we focused the presentation of our results on the degree of EEP determined from the velocity dispersion calculated in 3D spherical shells. In order to establish a closer contact with observations, now we study how the parameters and quantities discussed in the previous sections change when calculated on a 2D projection plane.
Analogously to the previous sections, we express time in our models in units of the initial half-mass relaxation time (here calculated from the projected half-mass radius, $R_\mathrm{h}$, and distinguished by an uppercase `R' in the subscript, i.e.~$t_\mathrm{Rh,0}$).
In addition to the equipartition mass, $\meq$, we also measure the evolution towards EEP by following the time evolution of the parameter $\eta$ which quantifies the dependence of the velocity dispersion on the stellar mass via $\disp \propto m^{\etam}$ (for systems at EEP, $\etam = -0.5$). Although $\meq$ provides a better description of the evolution towards EEP \citep[as shown in][]{bianchini_meq}, the parameter $\etam$ is still widely used in observational studies and, therefore, we include it in this section to ease the comparison with observations.

In Figs.~\ref{fig:meq_proj} \&~\ref{fig:eta_proj}, we plot the evolution of $\meq$ and $\etam$, respectively; in both isotropic and anisotropic models, with and without primordial binary stars, but only for the two extreme cases of the filling factor (i.e., 0.1 and 1.0). The three panels are again separated based on the Lagrangian radii of the shown percentage but we used the projected values this time. Hence, the same percentage in 2D corresponds to a smaller radius than in 3D, especially in the core where the line-of-sight contamination of the background and foreground stars is more relevant. With a star cluster placed in an external galactic potential, the kinematics of its outer regions may vary depending on the observer's position. In our models, the differences in $\meq$ (and $\etam$) calculated from the velocity dispersion projected on the $xy$, $xz$, $yz$ planes were only marginal and Figs.~\ref{fig:meq_proj} \&~\ref{fig:eta_proj} show their averages.

Although the projected tangential and radial components of the velocity dispersion are not equivalent to their 3D counterparts, we find the same trends identified and discussed above for the 3D calculations. The evolution of $\meq$ shows a similar behaviour in the central regions for both isotropic and anisotropic models and the kinematic trace of core collapse is clearly visible particularly in the models with no primordial binaries (see the top-left panel of both Figs.~\ref{fig:meq_proj} \&~\ref{fig:eta_proj}), and seems even more pronounced in projection than in 3D (compare Fig.~\ref{fig:meq_proj} with Figs.~\ref{fig:meq_iso} \&~\ref{fig:meq_aniso}).
In the regions around the half-mass radius, the projected values of $\meq$ also clearly show a more rapid evolution of the anisotropic systems towards EEP and that the rate of evolution towards EEP depends on the velocity dispersion component (see Fig.~\ref{fig:meq_proj_ratio}).
In the outer regions, the isotropic model still exhibits the initial evolution towards inverted EEP ($\meq{<}0$) followed by a transition to $\meq{>}0$ (i.e., the expected trend) late in the evolution, approximately around the time of core collapse.


As a final result, we show that all the trends identified with $\meq$ are still visible even if the degree of EEP is measured with the parameter $\etam$. In a cluster with stellar velocities independent of masses, $\etam$ would be zero, and as we noted, $\etam$ should decrease towards $-0.5$ when the cluster reaches EEP. As the cluster approaches core collapse, the inner regions show an upturn in $\etam$ (see the top-left panels of \ref{fig:eta_proj} \&~\ref{fig:eta_proj_radtan}); a similar feature is present in $\meq$ and, as discussed by \citet{bianchini_core_collapse}, is due to the combined effect of the radial variation of the velocity dispersion (increasing at smaller distances from the cluster's centre) and the spatial segregation of massive stars in the cluster central regions. In the outer regions of the models \texttt{i0.1s}, \texttt{i1.0s}, \texttt{i0.1b} and \texttt{i1.0b} (right-hand panels of Figs.~\ref{fig:eta_proj} \&~\ref{fig:eta_proj_radtan}), $\etam$ increases to positive values from the beginning of the simulations. This demonstrates that lower-mass stars acquire lower velocity dispersion than higher-mass stars over time. In the right-hand panels of Fig.~\ref{fig:eta_proj_radtan}, we can see that the tangential component is responsible for this inversion in the outer regions of the initially isotropic models, as only the values represented by the blue dashed lines are positive prior to core collapse (we note that the radial component of the \texttt{a0.1s} and \texttt{a0.1b} models also reach positive values before and during core collapse but these are merely fluctuations). As we already concluded with $\meq$, this result is independent of the primordial binary population and we do not see this behaviour in either of the initially anisotropic models.
Lastly, we note that while all the results are still visible when we use $\eta$ instead of $\meq$, the observational detection of the trends predicted by our analysis could be more difficult with this parameter.

\section{Conclusions}
\label{sec:concl}

In this work, we investigated the star cluster evolution towards energy equipartition (EEP). We focused on the effects of the external tidal field of the host galaxy, the role of a primordial binary population, and anisotropy in the initial velocity distribution. 
Here, we have studied only clusters on circular orbits but, in a future study, we plan to explore the dynamical effects of a time-varying potential for clusters on eccentric orbits.

Our numerical simulations show that the evolution towards EEP proceeds more rapidly in systems that were set up with radially anisotropic velocity distribution. Furthermore, in the intermediate and outer regions of both isotropic and anisotropic systems the rate of evolution towards EEP depends on the component of the velocity dispersion: in the initially anisotropic systems, the tangential component proceeds more rapidly towards EEP than the radial one, while the initially isotropic systems are characterised by the opposite trend.

We also find that in the outer regions, the initially isotropic systems evolve towards a state of `inverted' EEP in which low-mass stars have smaller velocity dispersion than high-mass stars, while the systems that started with the radial velocity anisotropy do not exhibit this behaviour. This result was first pointed out in \citetalias{pav_ves_letter} for a tidally underfilling star cluster with only single stars. In this work, we show that the tendency towards inverted EEP occurs regardless of the strength of the background tidal field and also independently of the population of primordial binary stars.
Our results further show that when we introduce a primordial population of binaries, the clusters tend to stay farther from EEP than the clusters in which no primordial binaries are present.

Finally, we find that our results also hold when the degree of EEP is measured with the 2D projected quantities that are usually calculated from observational data. Our findings can thus be tested and provide the theoretical tools to interpret existing \citep[e.g.][]{bellini_hst,bellini_hstII,libralato_hst,hst_legacy_xviii} and future observational studies of EEP in Galactic star clusters.

\section*{Acknowledgements}

This research was supported in part by Lilly Endowment, Inc., through its support for the Indiana University Pervasive Technology Institute. 
VP thanks Steven Shore for valuable discussion.
We are grateful to Antonio Sollima for comments that helped us improve the paper.

\section*{Data availability statement}

The data presented in this article may be shared on reasonable request to the corresponding author.



\bibliographystyle{mnras}
\bibliography{bibliography} 

%
%
%
%


\bsp	
\label{lastpage}
\end{document}